\documentstyle[12pt,epsfig]{article}
\def\v1{\vspace{1cm}}
\def\be{\begin{equation}}
\def\ee{\end{equation}}
\def\bc{\begin{center}}
\def\ec{\end{center}}

\def\vh{\varphi}

\newcommand{\bea}{\begin{eqnarray}}
\newcommand{\eea}{\end{eqnarray}}

\begin{document}
\setlength{\baselineskip}{5mm}
\hfill Preprint JINR E2-2002-149 \phantom{aaaaa}\\ [1cm]

\noindent{\large\bf
%------------------------------------------ Title --------------------
Origin of Matter from Vacuum in Conformal Cosmology
%---------------------------------------------------------------------
}\vspace{4mm}

\noindent{
%-------------------------------------- Author(s) --------------------
D.~Blaschke, V.~Pervushin, D.~Proskurin, S.~Vinitsky, and A.~Gusev
%---------------------------------------------------------------------
}\vspace{1mm}

\noindent{\small
%------------------------------------ Address(es) --------------------
Joint Institute for Nuclear Research, 141980 Dubna, Russia
%---------------------------------------------------------------------
}\vspace{4mm}

\begin{abstract}{
 We introduce the hypothesis that the matter content of the universe can be
a product of the decay of primordial vector bosons.
 The effect of the intensive
 cosmological creation of these primordial vector $W, ~Z $ bosons from the
vacuum is studied
 in the framework of General Relativity and the Standard Model
 where the relative standard of measurement  identifying conformal quantities
 with the measurable ones is accepted.
 The relative standard leads to
 the conformal cosmology with the z-history of
 masses with the constant temperature, instead of the conventional
z-history of the temperature with constant masses in inflationary cosmology.
 In conformal cosmology both
 the latest supernova data and primordial nucleosynthesis are compatible
 with a stiff equation of state associated with one of the possible states
of the infrared gravitation field.

 The distribution function of the created bosons
 in the lowest order of perturbation theory exposes a
 cosmological singularity as a consequence of the theorem about the
 absence of the massless limit of massive vector fields
 in quantum theory. This singularity
 can be removed by taking into account the collision processes leading to
a thermalization of the created particles.

The cosmic microwave background (CMB) temperature
$T=(M_W^2H_0)^{1/3}\sim 2.7 K$ occurs as an
integral of motion for the universe in the stiff state. We show that
this temperature can be attained by the CMB radiation
being the final product of the decay of primordial bosons.

 The effect of anomalous nonconservation of
baryon number  due to the polarization of the Dirac sea
vacuum by these primordial bosons is considered.
}
\end{abstract}

\section {Introduction}

In the inflationary models~\cite{linde} it is proposed that
 from the very beginning the universe is a hot fireball of massless particles
 that undergo a sequence of phase transitions. However, the origin of particles
 is an open question as the isotropic evolution of the
 universe cannot create massless particles \cite {tag, par, grib, sexl, zel, zel1}. Nowadays, it is evident that
 the problem of the cosmological creation of
 matter from the vacuum is beyond the scope of the inflationary model.

  The problem of cosmological particle creation in strong gravitational
 fields, in particular in the vicinity of the cosmological singularity,
 has been topical  for more than thirty years.
 For the first time this problem was described in~\cite {tag, par} and
 developed in~\cite {grib, sexl, zel, zel1}.

 Until now the consideration of the cosmological
 creation of particles was restricted only to scalar and spinor fields
 (though the necessity of including into consideration
 vector fields was discussed~\cite {grib}).
 It is considered that the number of created particles is obviously
 not sufficient for the description of primordial element abundance
 in the early universe~\cite {grib}.
 The problem of a possible origin of the matter in the universe as a result
 of the cosmological particle creation from vacuum is still under
consideration.

The investigation of the cosmological creation of massive vector bosons
from vacuum can introduce corrective amendments to the modern model of
the isotropic evolution of
the universe, since vector bosons detected at the CERN accelerator, are
 unique particles of the Standard Model (SM) which can exhibit
 a space singularity.

The matter is that the phenomenon of the cosmological creation
in isotropic models is described by the diagonalization of
equations of motion. This diagonalization
 includes a transformation into so-called conformal fields and coordinates.
In terms of conformal quantities the metrics is plane,
and the spatial scale factor is the scale of all masses, including the Planck mass
and masses of particles in the field theories, in particular in SM.
 For massive particles the cosmological singularity, i.e., disappearance of
 the spatial scale means disappearance of masses.
 The massive vector bosons are unique particles of SM
 which have a mass singularity~\cite{hp}.

 The absence of the massless limit of the massive Yang-Mills theory
 is a well-known fact~\cite{sf}. It leads, as we shall show below,
 to an ultra-violet divergence of the number of created longitudinal bosons
 calculated in the lowest order of perturbation theory.

 In the present paper, we consider the cosmological creation of
 massive vector bosons from vacuum in the conformal flat metrics,
 used for the description of the models of isotropic evolution of the
 universe.
 We investigate various possibilities to explain the origin of a
 visible matter as a finite product of decay of primary vector
 bosons, in agreement with the results of the primordial
 element abundance~\cite{three}
 and the latest Supernova data on the redshift - luminosity-distance
 relation~\cite{ps}.

 We list theoretical and observational arguments in  favour of the
 consideration of  created vector bosons
 as primary particles whose decay products form
  all visible matter of the universe,
 including CMBR and galaxies with baryon asymmetry.

 The present paper is an attempt to justify this statement,
 being grounded on the field theory, which unifies SM as the theory of
 fundamental particles in Riemannian space - time
 and general relativity (GR)
 with the relative measurement standard (the same problem was under consideration
 in a number of works~\cite{weyl,plb} in the framework of the conformal-invariant
 field theories).
 According to this standard,
 the extension of the universe assumes the extension of all intervals,
 including the measurement standards of these intervals.
 The relative measurement standard means identification of observable quantities
 with conformal fields and coordinates, which leads to the conformal
 cosmology~\cite{039} of the evolution of masses of the type of Hoyle-Narlikar~\cite{N},
 instead of the standard cosmology of distance evolution.
 The conformal cosmology~\cite{039} describes
 the present-day stage of accelerated evolution
 (according to the latest data on Supernova) and the stage  of primordial
 element abundance in the case of the stiff state
  which specifies the pure gravitational origin of dark
  energy~\cite{kasner, khal, tmf, 084}.

In the present paper, we used the
holomorphic representation of quantized fields~\cite{kar,ps1}
in terms of the creation and annihilation operators considered as
field variables (instead of stationary value
coefficients, as it was in the previous papers~\cite{tag} - \cite{zel1}).

 The contents of the paper is the following. The second section
 is devoted to the description of vector bosons and their cosmological creation.
 In the third section, the possibilities for substantiation of the CMB temperature
 in SM are investigated.
 In the fourth section, the baryon asymmetry of matter in the universe
 is considered.
In the conclusion, the arguments in favour of the Cold Scenario of the
evolution of the universe, based on the relative standard of measurement in GR,
are given.

%\vspace{6mm}
\section {Vector Bosons in the Isotropic Model of the Universe}

\subsection {Action}

 Consider "free" vector massive particles in  the homogeneous approximation
 of GR described by the Friedmann-Robertson-Walker cosmological models.
 Equations of motion of fields and metrics in the homogeneous approximation
 can be derived by the variation of the action
\be\label{tot}
S_{\rm tot}=S_{\rm univ}+S_{\rm v}~,
 \ee
where
\begin{equation}
S_{\rm  v}=\int d^4x \sqrt{-g}
\left[-{1\over 4}F_{\mu\nu}F^{\mu\nu}
-{1\over 2}M_v^2v_\mu v^\mu\right]~,~~~~~~
\label{Lem20}
\end{equation}
is the action of "free" vector $W^{\pm},Z^0$ bosons
${v_\mu}=({\bf v_0,v}_i)$ with the tension
 $F_{\mu\nu}=(\partial_\mu v_\nu-\partial_\nu v_\mu)$
 in the conformal-flat metric
  \be \label{cst}
  d{s}^2=g_{\mu\nu}dx^\mu dx^\nu
  = a(x^0)^2\left[(\bar{N}_0(x^0)dx^0)^2-(dx^i)^2\right],
 \ee
where $\bar{N}_0(x^0)$ is the lapse function defining the conformal time
\be  d\eta=\bar{N}_0(x^0)dx^0~; \label{cst1}
 \ee
\be\label{uncol1}
 S_{\rm univ}=- V_0
 \int\limits_{x^{0}_{1}}^{x^{0}_{2}} dx^{0}
 \left[ \vh_0^2\frac{(\partial_{0}a)^2}{\bar N_0}
 +\bar N_0\rho_{\rm univ}(a) \right],
\ee
is the action that gives the dynamics of the spatial scale factor
 under the supposition that the space is filled with the homogeneous
 "substance" with the conformal density $ \rho_{\rm univ} (a)$, to be
 defined below, $V_0 $ is a constant size of the coordinate space, and
\bea\nonumber
 \vh^2_0 = M^2_{\rm Planck} \frac{3}{8\pi} .
\eea
 We reiterated calculations of~\cite{hp} (where the action of vector fields
 is reduced into solutions of constraint equations, i.e.,
  equations on the time
 component) for the considered conformal flat metrics~(\ref {cst}),
 and rewrote the action~(\ref {tot}) in terms of the physical
 transverse and longitudinal variables
 \be \label{gradl}
S_{\rm tot}=\int\limits_{ }^{ }dx^0 \left\{ V_0\vh_0^2\left[-
 \frac{(\partial_0 a(x^0))^2}{\bar{N}_0}
 -\bar{N}_0 \rho_{\rm univ}(a)\right]
  + \bar{N}_{0}\int\limits_{V_0 } d^3x
{\cal L}_{\rm v}\right\}~,
\ee
 where ${\cal L}_{\rm v}={\cal L}^{\bot}_{\rm v}+ {\cal L}^{||}_{\rm v}$,
\begin{eqnarray}
{\cal L}_{\rm v}^{\bot} &=&
{1\over 2}\left[\frac{({\bf\partial}_0{\bf v}^{\bot})^{2}}{N_0^2}+
\left( {\bf v}^{\bot} \ , \ ({\bf\vec \partial}^2
- (M_va(x^0))^2){\bf v}^{\bot} \right)\right]
~,\nonumber\\
{\cal L}_{\rm v}^{||} &=& -{(M_va(x^0))^2\over 2}
\left[\left(\frac{{\bf\partial}_0{\bf v}^{||}}{{N_0}} \ , \ {1\over
{\bf\vec \partial}^2-(M_va(x^0))^2}
\frac{{\bf\partial}_0{\bf v}^{||}}{N_0}\right)+{\bf v}^{||2}\right]
\end{eqnarray}
are the Lagrangians of transverse and longitudinal bosons, and the brackets
 $({\bf v},{\bf w})$
designate a dot product of two vectors.

\subsection {Hamiltonian}

For definition of the law of evolution of all fields
 it is convenient to utillize the Hamiltonian form of action~(\ref {gradl})
in terms of  the Fourier-components
${\bf v}_k^{I}=\int\limits_{V_0}
d^3xe^{\imath\bf kx}{\bf v}^{I}({\bf x})$
\begin{eqnarray}
\label{grad}
S_{\rm tot}=\int\limits_{x^0_1 }^{x^0_2 }dx^0
\left( \sum\limits_{k}
 \left({\bf p}_{k}^{\bot}\partial_0{\bf v}_{k}^{\bot}
 + {\bf p}_{k}^{||}\partial_0{\bf v}_{k}^{||}\right)
-  P_{a} \partial_0a +\bar{N}_{0}\left[\frac{P_{a}^2}{4V_0\vh_0^2}-
V_0\rho_{\rm tot}(a)
\right]\right),
\end{eqnarray}
where
\begin{eqnarray}
\label{totgrad}
\rho_{\rm tot}(a)&=&\rho_{\rm univ}(a)+\rho_{\rm v}(a),~~~~\\
\rho_{\rm v}(a)  &=&V_0^{-1}(H^{\bot} + H^{||});
\label{totgrad1}
\end{eqnarray}
$H^{\bot}$ and $H^{{||}}$ represent Hamiltonians of free fields for
transverse and longitudinal components of vector bosons
\begin{eqnarray}
&H^{\bot} = \sum\limits_{k,\sigma} \frac{1}{2}\left[{\bf p}_k^{\bot}{}^2 +
\omega^2 {\bf v}_k^{\bot}{}^2\right]~,
\nonumber\\ [-8mm]&\label{grad1}
\\&\nonumber
H^{||} = \sum\limits_{k,\sigma} \frac{1}{2}\left
[\left(\frac{\omega(a,k)}{M_va}\right)^2{\bf p}_{k}^{||}{}^2 +
(M_va)^2 {\bf v}_{k}^{||}{}^2 \right]~,
\end{eqnarray}
with the dispersion relation in the form
 $\omega(a,k) = \sqrt{{\bf k^2} + (M_va)^2}$.

\subsection{Evolution of the Universe}

The basic equation of the FRW cosmology (i.e.,
the Einstein equation for the balance of energy densities)
is gained by variation of the action (\ref {tot}) with respect to the function
$\bar N_0$.

The variation of the action (\ref{tot})
determines the equations for the scale factor
\be\label{rho}
{\vh_0}^2[a'(\eta)]^2=
%\rho_c(\vh)+\rho_{\rm v}(\vh)\equiv
\rho_{\rm tot}(a)=
\rho_{\rm univ}(a)+
\rho_{\rm v}(a)\ee
and its solution
\be
\eta(a_I|a_0)=\vh_0\int\limits_{a_I}^{a_0}
{da\over\sqrt{\rho_{\rm tot}(a)}}
\ee
in terms of the conformal time $d\eta=\bar N_0 dx^0$, where $f'=df/d\eta$.

Transition to physical values (time $t_{\rm FRW}$, distance $l_{\rm FRW}$,
density $\rho_{\rm FRW}$) of the FRW cosmology is carried out with
the help of conformal transformations
\bea\label{preobr}
t_{\rm FRW}&=&\int_0^\eta d\bar\eta a(\bar\eta),\\
l_{\rm FRW}&=&a(\eta)r_{\rm},~~r_{\rm}=\sqrt{x_1^2+x_2^2+x_3^2}\\
\rho_{\rm FRW}(a)&=&\frac{\rho_{\rm tot}(a)}{a^4}\\
T_{\rm FRW}&=&\frac{T_{\rm CC}}{a}~.
\eea
In the terms of the FRW cosmology the equation of evolution takes
the conventional form
\be\label{preobrSK}
\vh_0^2H_{FRW}^2(t)=\rho_{\rm FRW}(a),
\ee
where $H_{FRW}(t)=a^{-1}(t)\dot{a}(t)$ is the Hubble parameter, and
 $\dot f=df/dt$.

Generally, the dependence of the conformal density of the universe
$\rho_{\rm univ} $, filled
with the homogeneous substance, on the scale factor $a(\eta)$
in the considered class of models for plane space
looks like
\bea\label{uncol}&
\rho_{\rm univ}(a)=\rho_{\rm Stiff}a^{-2}
+\rho_{\rm Rad}+\rho_{\rm M}a+\rho_{\Lambda}a^4.
\eea
Here $\rho_{\rm Stiff} $ describes the isotropic
contribution of the stiff state, for which
the density is equal to pressure\footnote{Nonisotropic version of this state with
subsequent isotropization \cite{grib,zel1}
was utilized for description of the early universe from 1921
\cite{kasner} to 1981; see, for example, \cite{grib,zel1,khal}.};
$ \rho_{\rm Rad}$ is utilized in the FRW cosmology for
the description of the radiation epoch of primordial nucleosynthesis;
$\rho_{\rm M}$ is the contribution of both visible and
invisible baryon substance;  $ \rho_{\Lambda}$ is the contribution vacuum of
Higgs scalar fields $ \rho_{\Lambda}$ in the inflationary epoch.

Solution of the equation of evolution of the scale with the initial data
$a (t_0) =1, ~ H _ {FRW} (t_0) =H_0 $ in terms of the world time $t$
for each equation of state looks like
\bea \label{SC}
a_{\rm Stiff}(t)&=&\sqrt[{\scriptstyle 3}~]{1 + 3\,H_0(t-t_0)},\\ \nonumber
a_{\rm Rad}(t)&=&\sqrt{1 + 2\,H_0(t-t_0)},\\ \nonumber
a_{\rm M}(t)&=&(1 + \frac{2}{3}H_0(t-t_0))^{3/2},\\ \nonumber
a_{\Lambda}(t)&=&e^{H_0(t-t_0)}~.
\eea
Rewrite these solutions
in  terms of the conformal time $\eta$ with the initial data
$a(\eta_0)=1,~~ a'(\eta_0)=da/d \eta=H_0 $
\bea \label{CC}
a_{\rm Stiff}(\eta)&=&\sqrt{1 + 2\,H_0(\eta-\eta_0)},\\ \nonumber
a_{\rm Rad}(\eta)&=&1 + H_0(\eta-\eta_0),\\ \nonumber
a_{\rm M}(\eta)&=&\left[1 + \frac{1}{2}H_0(\eta-\eta_0)\right]^{2},\\ \nonumber
a_{\Lambda}(\eta)&=&\left[{1-H_0(\eta-\eta_0)}\right]^{-1}~.
\eea
The equations on evolution of the scale
coincide with the equations on a collective motion of
spatial volume derived in GR~\cite{tmf,084}.

 The question arises: what is the reason of
 evolution of spatial volume (or the scale) of the universe?
 Till 1981 the predominant opinion was that such a reason is
 the matter of massive particles $ \rho _ {\rm M} (a) $,
  visible (baryons) and invisible
 (such as a massive neutrino), which before the recombination epoch
 ($z+1=1100 $) existed as radiation with the density
 $ \rho _ {\rm Rad} (a) $.
 The modern data on primordial nucleosynthesis and
  the element abundance in the universe convincingly testify to
  that the baryon matter is $30\div 20 $  times smaller
 than it is necessary for explanation of the evolution of the scale of
 the universe.

 The small contribution of the baryon matter to the evolution of the scale
 poses the problem on a source of this evolution in the long
 epoch of primordial nucleosynthesis
 (from $ 10^{-12} $ up to $ 10^{11} $ sec.). Its
  history testifies to the dependence of the scale on
 observable time as the square root.
 In the FRW cosmology this dependence~(\ref{SC}) is explained by the
  dominance
 of "radiation" with density $ \rho_{\rm Rad}(a) $.
 If the baryon matter is insufficient for evolution of the scale, then
 all known "radiations" are also insufficient for
 explanations of the evolution of the scale in the epoch of
 primordial nucleosynthesis.

 The newest data
 on the dependence of red shift on the distance up to Supernova \cite {ps},
  in the framework of the FRW cosmology, testifies to the
 dominance of the inflationary state of
 substance $ \rho_{\Lambda}(a) $.
 If one explains such a dominance by the "tail" of the epoch of
 primordial inflation
  of the early universe, one should also elucidate why there is
  an epoch with the dominance of "radiation"
 between two inflationary epoches (the primordial and the present-day)
  and where one should take
  so much "radiation" for visible evolution of the scale
 in this epoch of primordial nucleosynthesis.
 To answer these questions, we consider a wider supposition~\cite{039}
 that the evolution of the scale of the universe can be determined  not
 only by the theory and initial data, but also by the standard of measurement.

\subsection {Standard of measurement}

 It is worth reminding that the concept of measurable quantities
 in the field theory is no less important than
  the equations of the theory\footnote{''The most important aspect
 of any phenomenon from  mathematical
 point of view  is that of a measurable quantity.
  I shall therefore consider electrical phenomena
  chiefly with a  view to their measurement,
 describing the methods of measurement, and
 defining the  standards on
  which they depend.''
(J.C. Maxwell) \cite{JKMAXWELL}.}.

Suppose that nature selects itself both the theory and
standards of measurement, and the aim of observation is
to reveal not only initial data, but also these measurement standards.
In particular, one of the
central concepts of the modern cosmology is the concept of the scale defined
as a functional of spatial volume in GR~\cite{tmf,084}.
If expanding volume of the universe means
the expansion  of "all its lengths",
we should specify whether the measurement standard of length
expands.  Here there are two possibilities: the first, the absolute
  measurement standard does not expand;
  and the second, the relative measurement standard
 expands together with the universe.

Until the present time the first possibility was  mainly considered
in cosmology. The second possibility means that
we have no absolute instruments to measure absolute values
 in the universe. We can measure only a ratio of values which does not
 depend on the spatial scale factor. The relative measurement standard
 transforms the spatial scale of the intervals of lengths into
 the scale of masses which permanently grow.
The spectrum of photons emitted by atoms on far stars  two billion
years ago remembers a size of an atom which is determined by its mass. This
spectrum is compared with spectrum of similar atoms on the Earth
whose mass, at the present time, becomes much larger. This change of the mass
leads to red shift described by the relative (conformal) cosmology.
 This means that the observable time is identified with the conformal time, and
 the square root dependence  of the scale factor on the observable
 (i.e., conformal) time (\ref{evol}) follows from the stiff state.

 As it was shown in a recent paper~\cite{039},
 the stiff state in the framework of the conformal cosmology
  simultaneously describes the present-day stage of
accelerated evolution according to the latest data on Supernova.
These data together with the primordial nucleosynthesis can be treated
as  evidence of the relative standard
of measurement of the intervals of time and space in GR.

 In paper~\cite{084}, the stiff state in the
 conformal cosmology was described as
 a free motion of  metrics along a geodesic line in the "field space".
 The geometry of this "field space" in General Relativity
 was obtained  by Borisov and Ogievetsky~\cite{og}
 in terms of  the Cartan forms~\cite{car,dv} as a geometry of the coset
 of the affine group $A(4)$ over the Lorentz one $L$. The Cartan method
 of constructing   the nonlinear
 realization of the affine symmetry~\cite{car,dv}, in particular
 the operation of the group summation
 formulated in~\cite{vpc} -~\cite{isa}, allows us
 to introduce the concepts of "collective" and "relative"
 coordinates, and a class of "inertial motions" along
 "geodesic lines" into the coset $A(4)/L$ (i.e.,
    motions with constant canonical momenta).

Therefore,  considering the problem of creation  of vector bosons and
origin of the matter in the universe, we shall compare both
the measurement standards of the evolution.

\subsection {The evolution of vector bosons}

In the literature~\cite{tag}-\cite{zel1}, \cite{ps1}, the description of
the cosmological creation of particles in the considered class of models
includes transforming to conformal fields and coordinates
 \be \label{ccst}
 d{\bar s}^2=\bar g_{\mu\nu} dx^\mu dx^\nu= (\bar N_0 dx^0)^2-(dx^i)^2,
 \ee
 in  terms of which  the action for free
 vector W, Z  bosons takes the form
 \begin{equation}
S_{\rm v}=\int d^4x \bar{N}_0
\left[-{1\over 4}(\partial_\mu v_\nu-\partial_\nu v_\mu)
(\partial_{\mu'} v_{\nu'}-\partial_{\nu'} v_{\mu'})\bar g^{\mu\mu'}\bar g^{\nu\nu'}-
{1\over 2}a^{2}M_v^2v_\mu v_\mu\bar g^{\mu\nu}\right]~,
\label{Lem2}
\end{equation}
and the scale factor $a(x^0) $
becomes the scale of all masses, including  the Planck mass and
masses of particles (including
vector boson with mass $M_v $) in field theory.

Further calculations are conveniently carried out in terms of
"the running Planck mass" $\vh(x^0)\equiv\vh_0 a(x^0)$
with the  scale factor
\be \label{csm}
 M_va(x^0)=y_v\vh(x^0),~~~~~~~~~~~~~y_v=\frac{M_v}{\vh_0}~.
 \ee

In field theories ''particles'' are defined
as holomorphic  field variables
\bea \nonumber
   {\bf v}^{\bot}_k(x^0)\!&=&\!\sum\limits_{\sigma}
\frac{1}{
\sqrt{2V_0 \omega(\vh,k)}}
\bigl( a_{\sigma}^{\bot+}(-k,x^0){\bf e}_{\sigma}^{\bot}(-k)\!+\!
a_{\sigma}^{\bot}(k,x^0){\bf e}_{\sigma}^{\bot}(k)\bigr) \ ,\nonumber \\
%\eea
%\bea \nonumber
   {\bf p}^{\bot}_k(x^0)\!&=&-i\!\sum\limits_{\sigma}
\frac{\sqrt{ \omega(\vh,k)}}{\sqrt{2V_0 }}
\bigl( a_{\sigma}^{\bot+}(-k,x^0){\bf e}_{\sigma}^{\bot}(-k)\!-\!
a_{\sigma}^{\bot}(k,x^0){\bf e}_{\sigma}^{\bot}(k)\bigr) \ ,\nonumber \\
%\eea
%\bea \nonumber
 {\bf v}^{||}_k(x^0)\!&=&\!
 \frac{\sqrt{ \omega(\vh,k)}}{y_{v}\vh\sqrt{2V_0 }}
\bigl( a^{+}(-k,x^0)\!+\!
a^{||}(k,x^0)\bigr) \ ,\nonumber \\
 {\bf p}^{||}_k(x^0)\!&=&\! -i
 \frac{y_{v}\vh}{\sqrt{2V_0\omega(\vh,k) }}
\bigl( a^{||+}(-k,x^0)\!-\!
a^{||}(k,x^0)\bigr)~.
\eea
These variables are distinguished, since they diagonalize the
 field Hamiltonian
$$ H^I=
 \sum_{k,\sigma,I}^{ }\omega(\vh,k)({\hat N}_{k,\sigma}^I+ 1/2),
$$
where $\omega(\vh,k)=\sqrt{{\bf k}^2+y_v\vh^2}$ is the one-particle
energy,
 ${\hat N}^I =a_{\sigma}^{I+}(-k,x^0)a_{\sigma}^{I}(k,x^0)$
is the operator of the number of particles
with the spin $\sigma$ and $I=||,\bot $.
This definition of particles, by virtue of the dependence of
mass on time, leads  to nongiagonal terms in the canonical
 differential form, as  sources of the cosmological creation of particles
 \bea \nonumber
\left[\sum_{\varsigma}p_{\varsigma}^{I}\partial_0 {\rm v}_{\varsigma}^{I} \right]
&=&\sum_{\sigma,k}
\frac{\imath}{2}\left[ a^{{I}+}_{\sigma}(k,x^0){\partial_0 a^{I}}_{\sigma}(-k,x^0)
-a_{\sigma}^{I}(k,x^0){\partial_0 a}^{{I}+}_{\sigma} \right]
\\&&\nonumber- \sum_{\sigma,k}
\frac{\imath}{2}\left[a^{{I}+}_{\sigma}(k,x^0)a^{{I}+}_{\sigma}(-k,x^0) -
a_{\sigma}^{I}(k,x^0)a_{\sigma}^{I}(-k,x^0)\right]\partial_0 \Delta_{k}^{I}(\vh)~,
\eea
where $\varsigma$ is the index including momentum $k$ and
spin $\sigma$ of vector bosons,
\bea &\nonumber
\Delta^{{\bot}}_{v}(\vh) =
\frac{1}{2}\ln\left(\frac{\omega_{v}}{\omega_{I}}\right),~~~
\\&\nonumber
\Delta^{||}_{v}(\vh) =\ln\left(\frac{\vh}{\vh_{I}}\right)
- \frac{1}{2}\ln\left(\frac{\omega_{v}}{\omega_{I}}\right)~,
\eea
and  $\vh_I$ and  $\omega_I$ are cosmic initial data.
%beginnew
The classical equations in terms of "particles" can be written as
\be
i\frac{d}{d\eta}\chi_{a_{\varsigma}}=-{\hat H}_{a_{\varsigma}}\chi_{a_{\varsigma}}.
\ee
 where
\be
\bar\chi{}_{a_{\varsigma}}=(a_{\varsigma},-a_{\varsigma}^+);\;\;\chi_{a_{\varsigma}}=\left(
\begin{array}{c} a_{\varsigma}^+\\a_{\varsigma} \end{array}\right);\;\;{\hat H}_{a_{\varsigma}}=\left|
\begin{array}{ccc}\omega_{a_{\varsigma}} &,&-i\Delta_{\varsigma}\\ \\ -i\Delta_{\varsigma}&,&-\omega_{a_f}\end{array}
\right|.
\ee
After the Bogoliubov transformations
\bea &
b_{\varsigma}^+=\alpha^*_{\varsigma}a_{\varsigma}^{+}
 +\beta^*_{\varsigma}a_{\varsigma}
\nonumber\\ [-8mm]&\label{bogtr}
\\&\nonumber
b_{\varsigma}=\alpha_{\varsigma}a_{\varsigma}
 + \beta_{\varsigma}a_{\varsigma}^{+}
\eea
written briefly as
\be
\chi_b=\hat O\chi_a;\;\;\;
\hat O= \left(
\begin{array}{ll} \alpha^*,&\beta^*\\ \\
\beta,&\alpha\end{array}\right);\;\;\; \hat O{}^{-1}= \left(
\begin{array}{rr} \alpha,&-\beta^*\\ \\ -\beta,&\alpha^*\end{array}\right);
\ee
this equation becomes
\be
i\frac{d}{d\eta}\chi_b=[-i\hat O{}^{-1}\frac{d}{d\eta}\hat O -
\hat O{}^{-1}\hat H{}_a\hat O]\chi_b\equiv -\hat H{}_b\chi_b.
\ee
Let us require  $\hat H{}_b$ to be diagonal
\be
\hat H{}_b=\left(
\begin{array}{rr} \omega_b,&0\\ \\ 0,&-\omega_b\end{array}\right).
\ee
This means that $\alpha$ and $\beta$  satisfy the equations
\be
\omega_b=(|\alpha|^2+|\beta|^2)\omega_a-i(\beta^*\alpha-\beta\alpha^*)\Delta-
i(\beta^*\partial_T\beta-\alpha\partial_T\alpha^*),
\ee
\be
0=2\beta\alpha\omega_a-i(\alpha^{2}-\beta^{2})\Delta-
i(\alpha\partial_T\beta-\beta\partial_T\alpha).
\ee
For
\be
\alpha=\cosh(r)e^{i \theta}\;;\;\;\;\beta=i\sinh(r)e^{-i \theta}
\ee
these equations convert into
 the equations for the coefficients of the Bogoliubov transformation
as conditions of  diagonalization ~\cite {ps1}
\bea\label{main}
[\omega_{\varsigma} - \theta'_{\varsigma}] \sinh(2r_{\varsigma})&=&
\Delta'_{\varsigma}\cos(2\theta_{\varsigma})\cosh(2r_{\varsigma})~, \\
r'_{\varsigma} &=& - \Delta'_{\varsigma}\sin(2\theta_{\varsigma})~.\nonumber
\eea
These coefficients determine the number of  particles
\be
{\cal N}_{\varsigma}(\eta) =
{}_{\rm sq} \langle 0|\hat N_{\varsigma}|0\rangle_{\rm sq}
= \sinh^2 r_{\varsigma}(\eta)\nonumber
\ee
created during the time $ \eta $ from "squeezed" vacuum defined as
$$
b_{\varsigma}|0\rangle_{\rm sq} = 0~.
$$
The density of created particles is
\be\label{v}
\rho_{\rm v} = \sum_{\varsigma} \omega_{\varsigma}(\vh)
{}_{\rm sq}(\langle 0|\hat N_{\varsigma}|0\rangle_{\rm sq}+ 1/2)~.
\ee
Further, we restrict ourselves to the cases when the back reaction of
  this density on the evolution of the universe can be neglected.

\subsection {Initial data}

To find initial data for the Bogoliubov equation~(\ref {main}),
we make a change of variables
$(r_{\varsigma},\theta_{\varsigma}\rightarrow C_{\varsigma},{\cal N}_{\varsigma})$
\bea
\label{sub}
\cos(2\theta_{\varsigma})\sinh(2r_{\varsigma})&=&C_{\varsigma}~,\nonumber\\
\sinh(2r_{\varsigma})&=&\sqrt{{\cal N}_{\varsigma}({\cal N}_{\varsigma}+1)}~.
\eea
Then, equations~(\ref{main}) become
\bea \label{3}
{\cal N}'_{\varsigma}&=&
\left( \frac{\Delta'_{\varsigma}}{2\omega_{\varsigma}} \right)C_{\varsigma}'
~,\nonumber\\
{\cal N}'_{\varsigma}&=&
-\Delta'_{\varsigma}\sqrt{ 4 {\cal N}_{\varsigma}({\cal N}_{\varsigma}+1)
- C_{\varsigma}^2}~.
\eea
From equations~(\ref{3}) it follows that {\it vacuum}
initial states ${\cal N}_{\varsigma} (\eta=0) =0 $
correspond to the value $C_{\varsigma} (\eta=0) =0 $.
This gives the following initial values of the coefficients $r$ and $\theta$
\be \label{id}
r_{\varsigma}(\eta=0)=0,~~~~~~~~~~~~~~~~
{\theta}_{\varsigma}(\eta=0)=\frac{\pi}{4}~.
\ee

\subsection {The distribution function }

Let us consider an example of the solution of the obtained set of
equations for
the stiff state $ \rho =\rho_{\rm Stiff} $ neglecting
the back reaction of the density of created bosons on the evolution of the
universe.

In this case, the Bogoliubov equations~(\ref {main}),
rewritten in the dimensionless variables
\be\label{g}
 \tau=\eta 2H_I=\eta/\eta_I,~~~~~~~~~~~ x=\frac{q}{M_{I}}
\ee
and initial data
$M_{I}=M_v(\eta=0),~~~H_I=H(\eta=0)$ with taking into account
the dispersion relation  $\omega_v=H_I\gamma_v\sqrt{1+\tau+x^2}$, where
\be\label{gv}
\gamma_v=\frac{M_{I}}{H_I}~,
\ee
take the form
$$
\left[\frac{\gamma_v}{2}\sqrt{(1+\tau)+x^2} -
\frac{d\theta^{||}_{v}}{d\tau}\right] \tanh(2r^{||}_{v}) =
\frac{1}{4}\left[\frac{2}{(1+\tau)}-\frac{1}{\left[ (1+\tau)+x^2\right]}\right]
\cos(2\theta^{||}_{v})~,
$$
$$
\frac{d}{d\tau}r^{||}_{v} =
-
\frac{1}{2}\left[\frac{1}{(1+\tau)}-\frac{1}{2}\frac{1}{\left[ (1+\tau)+x^2\right]}\right]
\sin(2\theta^{||}_{v})~,
$$
$$
\left[\frac{\gamma_v}{2}\sqrt{(1+\tau)+x^2} -
\frac{d}{d\tau}\theta^{\bot}_{v}\right]  \tanh(2r^{\bot}_{v}) =
\frac{1}{4}\left[\frac{1}{(1+\tau)+x^2}\right]
\cos(2\theta^{\bot}_{v})~,
$$
$$
\frac{d}{d\tau}r^{\bot}_{v} =
-\frac{1}{4}\left[\frac{1}{(1+\tau)+x^2}\right]
\sin(2\theta^{\bot}_{v})~.
$$
These equations were solved numerically at positive values of momentum
$x=q/M_I $, by
utillizing the asymptotics of the solutions $r (\tau) \to {\rm const} \cdot\tau $,
$ \theta (\tau) = \pi/4+O (\tau) $ from the neighborhood $ \tau=0 $.
The distribution functions of the longitudinal
$ {\cal N} ^ {||} (x, \tau) $ and
transverse $ {\cal N} ^ {\bot} (x, \tau) $ vector bosons
for the initial data $H_I=M_I,~\gamma_v=1 $,
are introduced in Fig. 1.
\begin{figure}[t]
 \includegraphics[width=0.57\textwidth,height=0.51\textwidth,angle=-90]{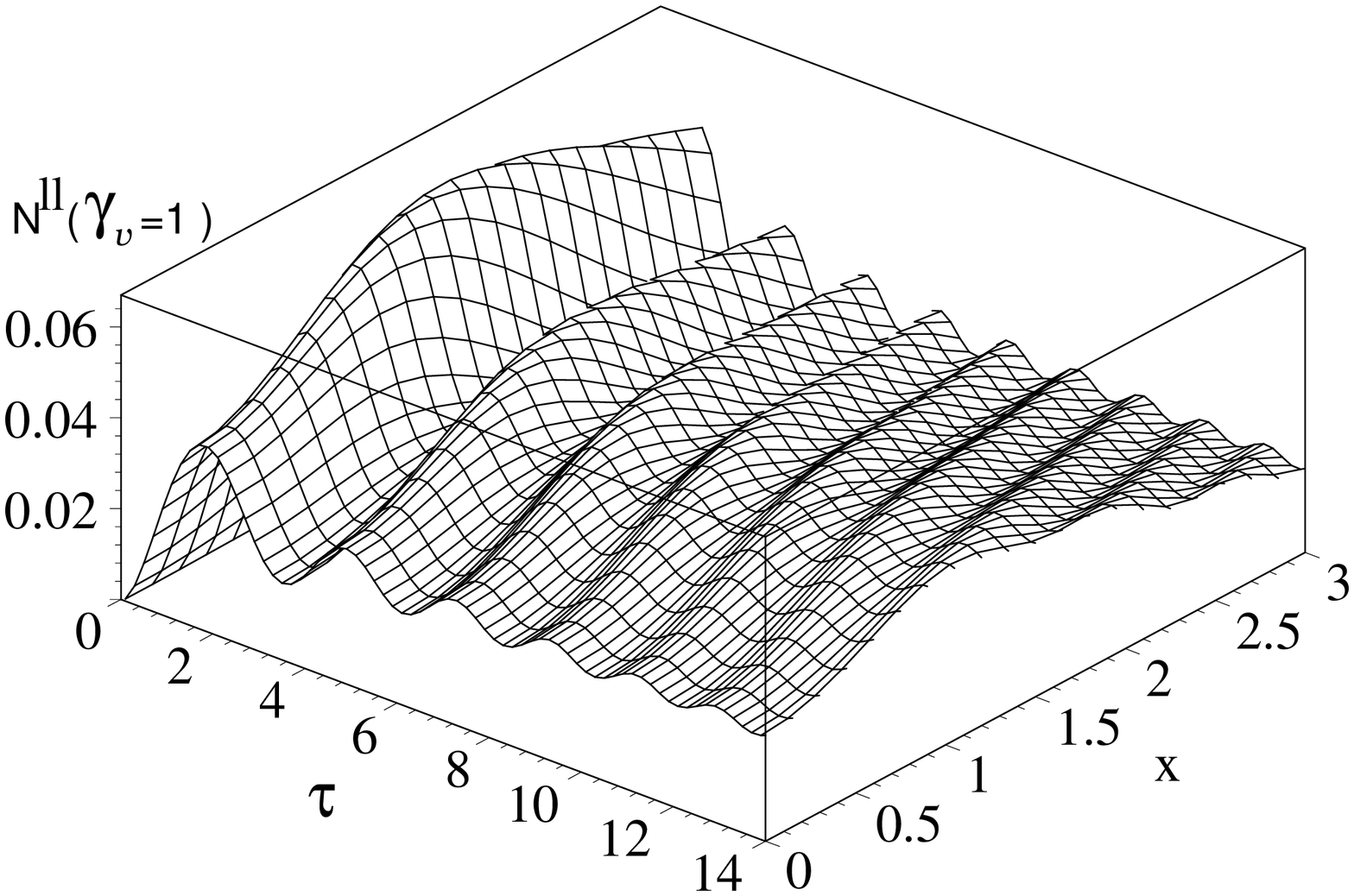}\hspace{-5mm}
 \includegraphics[width=0.57\textwidth,height=0.51\textwidth,angle=-90]{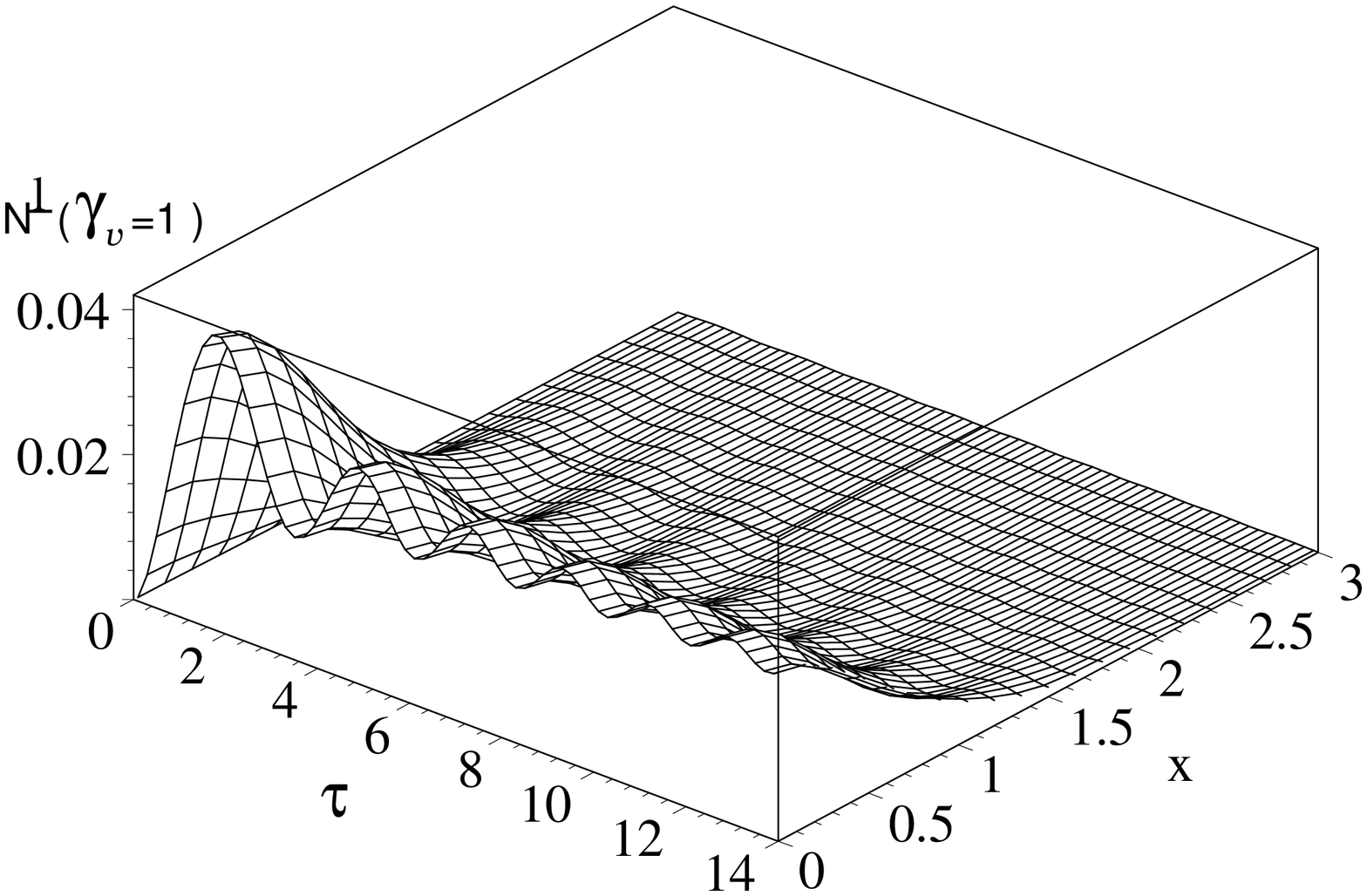}
  \caption{ Dependence of
longitudinal $N^\|$ and transverse $N^\bot $  components of the
distribution function of vector
bosons from dimensionless time $ \tau=2H_I \eta $ and dimensionless
momentum~ $ x = q/M_I $ at
value of the initial data $M_I = H_I$ ($\gamma_v = 1$).}
\end{figure}

The choice of these initial data is
determined by the lower
 boundary for a boson mass from the area of its initial values
 allowed by the uncertainty  principle $ \delta E ~\eta_I\geq \hbar $
for energy variations of energy $ \delta E=2M_I $
at creation of a pair of bosons in the universe with
 minimum lifetime for a considered case $ \eta_I=1/2H_I $.

From Fig. 1. one can see that a longitudinal
component of the distribution function
 is essentially larger than the transverse one, which
 demonstrates a more intensive cosmological creation of
longitudinal bosons in contrast with transverse ones.
The sluggish decrease with momentum of
longitudinal components is explained
by mass singularity of a distribution function of
longitudinal vector bosons~\cite{hp,sf}.
One of the consequences of such a decrease is the divergence
 of the density of created particles~\cite{par}
\be\label{nb}
 n_{v}(\eta)=\frac{1}{2\pi^2}
\int\limits_{0 }^{\infty }
dq q^2
\left[ {\cal N}^{||}(q,\eta) + 2{\cal N}^{\bot}(q,\eta)\right]\to\infty~.
\ee
It is possible to test that the divergence takes place for any
equation of state.
That is the shortage of the lowest order of  perturbation theory,
where one neglects interactions of vector bosons,
including the scattering processes forming an integral of
collisions in the kinetic equation for distribution functions.

 The divergence of the integral over momentum and the mass
 singularity  give a possibility of
 explaining  the genesis of the matter in the universe
 and its temperature by cosmological creation of particles from vacuum.
 Let us remind that the problem of  genesis of the primordial
 particles is not considered  in the inflationary model of the universe.
 This model is based on the conjectures \\
1) nonzero vacuum density of a scalar Higgs field , \\
2) nonzero initial data for numbers of all particles
considered as massless ones, \\
3) dependence of a homogeneous scalar field on temperature,
the reason of forming of which is obscure, \\
4) the concept of temperature in a strongly
nonequilibrium regime of  the inflationary expansion.

According to the standard cosmological model of the expanding hot universe,
the existence of the temperature of matter removes
the divergence of the integral over momenta by the corresponding Boltzmann
factor. In this case, at a value of temperature of
about boson mass, the universe at the stage of the radiation dominance
is capable of creating so many bosons with the density of
about $T^3 $, that they can explain the very
stage of radiation dominance, if the radiation is considered as the final
 product of decays of primordial bosons. Let us remind
that the radiation dominance in the standard cosmological model
 is  the condition for the element abundance and the chemical evolution
 of the matter.
This result poses the following problems:
What was in the expanding universe before creation of vector bosons?
 Is it possible to explain all
observed matter (with CMB radiation and baryon-asymmetric matter) by
 its cosmological creation from vacuum at the radiation dominance stage?

The latest data on  the dependence of the redshift - distance-relation
 on distance to Supernova~\cite {ps} testifying
to the dominance in the universe of the unobservable dark energy
(called  Quintessence \cite {dark}) have posed the problem of
explanation, in the Hot Universe Scenario, of
 the appearance of a pure radiation stage  between two
  inflationary stages - rapid (primordial) and sluggish
 {\it Quintessence},  both have no a bearing on
   the observed matter?
 Why does the inflationary stage  appear again after the  radiation epoch
  as it follows from the latest  Supernova data~\cite{ps}?
  All these problems have not yet found a satisfactory explanation
 within the framework of the inflationary model and the Hot Universe Scenario.

\section {The temperature of CMB Radiation}

\subsection {The relative standard of measurement of intervals}

For  solution of the problem of the genesis of the matter in the universe
in the context of the obtained effect of intensive creation of vector bosons
we consider the  Cold Universe Scenario based on the relative
 measurement standard in GR.

 The fixed fact of the primordial
 nucleosynthesis \cite{three} testifies only to that the scale factor
 is proportional to the square root of the observed time
 \be\label{evol}
 a^2(\eta)=a_I^2(1+2H_I\eta)=(1+2H_0(\eta-\eta_0)),
 \ee
 where $a_I$ and $H_I$ are primordial values of the cosmological scale and
 the Hubble parameter, and $a_0=1$, $H_0$ are their present-day values connected
 with the first by the integral of motion
 \be\label{int}
 a_I^2H_I=a^2(\eta)H(\eta)=H_0~.
 \ee

If we identify the observed time with the conformal one,
then the proportionality of the
scale factor to the square root of the observed (i.e., conformal)
time (\ref{evol}) follows  from the stiff  state
where the density of energy is equal to that of pressure.
 As it has been shown in a recent paper~\cite{039}, the stiff state is compatible
with the latest Supernova data
on the accelerating universe evolution~\cite{ps}.
Moreover, the conformal
version of the stiff state reproduces the z-history of the chemical
evolution of the element abundance in the FRW cosmology,
since we have, in the conformal cosmology, the same square root
dependence of the scale factor on the observable time in the stiff
stage. Therefore, we can utillize the rigid equation of state
 for  estimation of temperature of the CMB
  radiation as a final product of the decays of primordial
 vector bosons.

The primordial value $H_I $ of the Hubble parameter
sets a natural unit of time $ \eta_I=1/2H_I $ as a minimal
lifetime of the universe. As it follows from the uncertainty principle for
energy $\delta E ~\eta \geq \hbar$, a
characteristic time of all physical processes with variation
of energy $ \delta E $  is less than this lifetime $ \eta_I = 1 / 2H_I $.
We can speak about the cosmological creation of a pair
of  massive particles in the universe,  when the particle mass
$ M_v(\eta=0)=M_{I}$  is larger than the primordial Hubble parameter
$M_{I} \geq H_I $. Therefore, it is worth introducing a ratio of the
initial data $M_{I}/ H_I=\gamma_v\geq 1$ (\ref{g}).

\subsection {The estimation of the
temperature of the CMB  Radiation}

To remove the divergence in the integral (\ref{nb})
and  estimate the temperature of the CMB radiation,
we multiply the distribution function of the primordial bosons
${\cal N}^{||}(q,\eta)$ and ${\cal N}^{\bot}(q,\eta)$
by the Bose-Einstein distribution function (with the Boltzmann constant
$k_{\rm B}=1$)
\be \label{bose}
{\cal F}\left(T,q,M_v(\eta),\eta\right)=
\left\{\exp\left[\frac{\omega_v(\eta)- M_v(\eta)}{T}\right]
-1\right\}^{-1}~,
\ee
where $T$ represents the cutoff parameter. In this case, the
expression for density (\ref{nb}) takes the form
\be\label{n1}
 n_{v}(T,\eta)=\frac{1}{2\pi^2}
\int\limits_{0 }^{\infty }
dq q^2{\cal F}\left(T,q,M(\eta),\eta\right)
\left[ {\cal N}^{||}(q,\eta) + 2{\cal N}^{\bot}(q,\eta)\right]~.
\ee
Let us propose that a characteristic duration of all processes in the universe
 should not exceed the minimal lifetime of the universe, we
 can  estimate values of integrals~(\ref{n1}) under the condition
 that initial data for the temperature
 inherits the initial data for the Hubble parameter
 \be \label{tcmbr}
 T= M_{I}= H_I~.
 \ee
 In this conjecture the temperature as a constant of the conformal cosmology
 is the single integral of the   stiff state~(\ref{int})
\be \label{tcmbr1}
 T=\left[M^{2}_{I}H_I\right]^{1/3}
 =\left[M^{2}(0)H_0\right]^{1/3} = 2.76~{\rm K}~,
\ee
 where $M(0)$ coincides with the present-day value of the mass of
 the vector boson $M_W$.

 Now we  show that such a coincidence with the CMB temperature
 is not accidental. Calculations demonstrate rapid
 determination of the density of vector bosons~(\ref{n1})
 (during the time $\eta_I=1/2H_I$) in an equilibrium  state
 where the weak dependence of the density on time is observed.

 The dominating contribution of large momenta to the
 integral (\ref{n1}) (see Fig.1.) means the relativistic
 dependence of the density on temperature
 $$
 n_{v} = C T^3~.
 $$

The numerical calculation of the integral~(\ref {n1}) gives for
the constant $C $ the following value:
\be \label{nc}
C = \frac{ n_{v}}{ T^3} = \frac{1}{2\pi^2}
\left\{ [1.877]^{||}+2 [0.277]^{\bot}=2.432   \right\}~,
\ee
where the contributions of longitudinal and transverse bosons are indicated
by the subscripts $(||,~ \bot)$.

The thermal equilibrium is formed, if the relaxation time of the process
of establishment of temperature of vector bosons
\be \label{rel01}
\eta_{\rm relax.} = \frac{1}{n(T^3)\sigma_{\rm scat.}} \le \frac{1}{2H_I}
\ee
is less than the time for creating vector bosons. Fig. 1 shows us
that the latter is the order of
 lifetime of the Universe $\eta_I=1/2 H_I$. The condition (\ref{rel01})
  is fulfilled for
\be \label{rel02}
\sigma_{\rm scat.} = \gamma_{\rm  scat} / M_{I}^2~,
\ee
if $C \gamma_{\rm  scat}>2$.

 On the other hand,  it is possible to estimate the
 lifetime of the created bosons in the early universe
 in dimensionless unities
$ \tau_L =\eta_L/\eta_I $, where $ \eta_I = (2H_I) ^ {-1} $, by
utillizing an equation of state $a^2 (\eta) =a_I^2 (1 +\tau_L) $ and
define the lifetime of $W$-bosons in the Standard Model
\be \label{life}
1+\tau_L=
\frac{2H_I\sin^2 \theta_{\alpha_{\rm QED}\rm W}}{ M_W(\eta_L)}=
\frac{\sin^2 \theta_{\rm W}}{\alpha_{\rm QED}\gamma_v\sqrt{1+\tau_L}}~,
\ee
where $\theta_{\rm W}$ is the Weinberg angle,  $\alpha_{\rm QED}=1/137$.

The solution of equation~(\ref{life})
\be \label{lifes}
\tau_L+1=
\left(\frac{2\sin^2\theta_{\rm W}}{\gamma_v\alpha_{\rm QED}}\right)^{2/3}
\simeq \frac{16}{\gamma_v^{2/3}}~
\ee
gives an estimation of the lifetime of the created bosons for $\gamma_v=1$
\be \label{lv}
\tau_L =\frac{\eta_L}{\eta_I}\simeq \frac{16}{\gamma_v^{2/3}}-1=15~,
\ee
which is 15 times larger than the relaxation time.

Therefore,  the relaxation time is much less than the
 lifetime of vector
bosons, and we can introduce the concept of the temperature of vector bosons,
which is inherited by final products of their decays,
i.e. gamma-quanta, forming,
according to the modern point of view, the CMB radiation in the universe.
It is easy to show that the temperature of photons $n_\gamma\simeq n_v $
coincides with the constant temperature
of the primordial bosons $T\simeq 2.76 K $.

Really,
if one photon goes from annihilation of products of the decay of
$W^{\pm}$  bosons,
and another photon - of $Z $ bosons, we get the density of photons
with a constant temperature
\be \label{1nce}
\frac{ n_{\gamma}}{ T^3}=\frac{1}{\pi^2}
\left\{ 2.432  \right\}~.
\ee
This temperature is of an order of
the temperature of the CMB radiation $ T=T_{\rm CMB}= 2.73~{\rm K}$.

The temperature of photon radiation, which appears after annihilation and
decays of $W^{\pm}$ and $Z $ bosons in the conformal cosmology, is invariant
and the simple estimation fulfilled above gives the value surprisingly close
to the observed temperature of the CMB radiation which is
determined in the conformal cosmology as a
fundamental constant - the integral of the stiff
state~(\ref{tcmbr1}).

\subsection {The back reaction of created particles on the evolution of the Universe}

The equation of motion
 $ \vh'^2(\eta)= \rho_{\rm tot}(\eta)$ with the Hubble parameter
 defined as $ H =\vh'/\vh $  means that
the energy density of the universe at any moment is equal to the
 so-called critical density
$$
 \rho_{\rm tot}(\eta)= H^2(\eta)\vh^2(\eta)
\equiv \rho_{\rm cr.}(\eta)~.
$$
The permanent dominance of the matter with
the stiff state
means the existence of the integral of motion
$$ H(\eta)\vh^2(\eta)=H_0\vh_0^2~.
$$
Now find a ratio of the density
of the created matter
$ \rho_{\rm v}(\eta_I)\sim  T^4\sim H_I^4\sim M^4_{I}$ to the
density of the primordial cosmological motion of the universe
$ \rho_{\rm cr.}(\eta)=H_I^2\vh^2_{I}$. This ratio has an
extremely small number
$$\frac{\rho_{\rm v}(\eta_I)}{\rho_{\rm cr.}(\eta_I)}
=\frac{M^2_{I}}{\vh_I^2}=\frac{M^2_{W}}{\vh_0^2}=y^2_v=10^{-34}.$$
Therefore the back reaction of created particles
on the evolution of the universe is a negligible quantity.
For the lifetime of the universe
the primordial density of the cosmological motion
$ \rho_{\rm tot}(\eta)=H_I^2\vh^4_{I}/\vh^2(\eta)
\equiv H_0^2\vh^4_{0}/\vh^2(\eta)$
 decreases by  $10^{29}$ times, and in the present-day
 epoch the critical density
$\rho_{cr \, 0}\equiv H_0^2\vh^2_{0}=10^{-29}\rho_{cr \, I}$ is
$20\div30 $ times greater than the density of the observed baryon matter.\\

\section {Baryon Asymmetry of Matter in the Universe}

The baryon asymmetry of the matter in the universe can be formed
by polarization  of the Dirac sea vacuum by fundamental
bosons during their life.

Interaction of the primordial $W $ and $Z $ bosons with the
left-hand fermions leads to nonconservation of fermion quantum numbers.
It is known that the gauge-invariant current of each doublet is saved
only at a classic level~\cite{ufn}.

At a quantum level, we have an abnormal current
$j_L^{(i)}=\psi^{(i)}_L\gamma_\mu\psi^{(i)}_L$,
\bea \nonumber
\partial_\mu j_L^{(i)}=-\frac{{\rm
Tr}F_{\mu\nu}{}^*\!{F_{\mu\nu}}}{16\pi^2}, ~~
F_{\mu\nu}=-\frac{\imath g\tau_a}{2}{F_{\mu\nu}^{a}}^{\rm(p.t.)}, ~~
{F_{\mu\nu}^{a}}^{\rm(p.t.)}
=\partial_\mu v_\nu^a-\partial_\nu v_\mu^a~.\eea

During their lifetime, the vector bosons polarize  the Dirac sea of
the left-hand fermions with negative energy.
The number of the left-hand fermions $N (\eta_L) $
is determined by the expectation value of the Chern-Simons functional
from the Bogoliubov vacuum~\cite{ufn}
\bea \nonumber N(\eta_L)=
 -\int_0^{\eta_L} d\eta \int \frac{d^3 x}{16\pi^2} \;
 {}_{\rm sq}\langle 0|{\rm Tr}F_{\mu\nu}
 {}^*\!{F_{\mu\nu}}|0\rangle{}_{\rm sq} ,
\eea
 where $\tau_L =\eta_L/\eta_I$ is the lifetime of bosons.
During their lifetime, transverse vector bosons are evolving so
that the Chern-Simons functional is changed. If we take the integral over
four-dimensional conformal
space-time confined between three-dimensional hyperplanes
$\eta=0$ and $\eta=\eta_L$, we find that the number of left fermions
$N(\eta_L)$ is equal to the  Chern-Simons functional
 \be
\label{nw}
\Delta N_W =
\frac{4{\alpha}_{\rm QED}}{\sin^{2}\theta_{\rm W}}\int_{0}^{\eta^{W}_{\rm l}}
d\eta \int \frac{d^3 x}{4\pi}~~{}_{\rm sq}
\langle 0|E^{W}_{i}B^{W}_{i}|0\rangle{}_{\rm sq}~,
\ee
where $E_i$ and $B_i$ are the electric and magnetic fields strengths.
The squeezed vacuum and Bogoliubov transformations (\ref{bogtr}) give
a nonzero value for these quantities
\be
\label{ev}
\int  \frac{d^3 x}{4\pi}~{}_{\rm sq}
\langle 0|E^{v}_{i}B^{v}_{i}|0\rangle{}_{\rm sq}
= -\frac{V_0}{2} \int\limits_{0 }^{\infty }dk |k|^3
\cos(2\theta_\zeta)\sinh(2r_\zeta)~,
\ee
where $\theta_\zeta$ and $r_\zeta$ are given by  equation (\ref{main})
 for transverse bosons.
 Using  the relation
\be
\label{nz}
\Delta N_Z =
\frac{{\alpha}_{\rm QED}}{\sin^{2}\theta_{\rm W}\cos^{2}\theta_{\rm W}}
\int_{0}^{\eta^{\rm Z}_{\rm l}} d\eta \int
\frac{d^3 x}{4\pi}~~{}_{\rm sq}
\langle 0|E^{Z}_{i}B^{Z}_{i}|0\rangle{}_{\rm sq}~,
\ee
we find the Chern-Simons functional for Z bosons in a similar way.

If we take into account the numerical evaluation of the integral (\ref{ev}), in
the conjecture that the lifetime of bosons is
$ \tau_L^W = 15 $, $ \tau_L^Z = 30 $,
 we estimate the magnitude of the nonconservation of the fermion number
\begin{eqnarray}
\Delta F&=&\frac{(\Delta N_W+\Delta N_Z)}{V_0}
\nonumber\\
 &=&\frac{{\alpha}_{\rm QED}}{\sin^{2}\theta_{\rm W}}
  \left(\frac{ T^3 2.402}{\pi^2}\right)
  \left(4\times 1.44+\frac{2.41}{\cos^{2}\theta_{\rm W}}\right)
  \left(\frac{\pi^2 }{2.402}\right)=1.2  n_{\gamma}~,
\end{eqnarray}
where $n_{\gamma}$ is the density of the number of the CMB photons(\ref{1nce}).
The baryon asymmetry appears as a consequence of three Sakharov
conditions~\cite{ufn}: the
${\rm CP}$-nonconservation, the evolution of the  universe $H_0\not= 0$
and the violation of the baryon number
$$
{\Delta B}=X_{\rm CP}\frac{\Delta F}{3}=0.4X_{\rm CP}n_{\gamma}~,
$$
where $X_{CP}$ is a factor determined by a superweak interaction
of $d$ and $s$-quarks $(d+s~\rightarrow ~s+d)$
with the CP-violation experimentally observed in decays of
$K$ mesons~\cite{o}.
 From a ratio of the number of baryons to the number of
  photons it is possible to make an estimation of
  magnitudes of a constant of a weak coupling
   $X_{\rm CP}\sim 10^{-8}$.\\

\section {Conclusion}

The main result of the paper is the description of the effect of the
intensive creation of longitudinal vector bosons from vacuum with
the divergent integral for the density of the number of created particles
in the lowest order of perturbation theory.
The divergence of the integral is a corollary that the
massless limit of the massive vector theory does not exist (as it was
revealed as far back as early papers by Ogievetsky, Polubarinov
as well as  Faddeev and Slavnov).
The back reaction of created particles
on the evolution of the universe is a negligible quantity.

We have considered this effect, comparing  two possible
measurement standards: the absolute (not extending together with the universe)
and the relative (extending together with the universe), appropriate to
the Friedmann-Robertson-Walker cosmology and the conformal cosmology.
These standards are connected by the conformal transformations.

In the case of the Friedmann-Robertson-Walker cosmology of the expanding
hot universe the primordial temperature cuts the divergent
integral by the Boltzmann factor.

At the value of the temperature about the mass of bosons, the universe
at the stage of the radiation dominance is able of creating
 many bosons
with the density of the order of $T^3$.
The question arises: how many bosons are necessary for an explanation
of the very stage of the radiation dominance,
if one considers the radiation as the final product
of the decay of primordial bosons.

This result poses problems which up till now have no answer
in the FRW cosmology. What is the origin of the
primordial temperature?
Is it possible to introduce the concept of temperature into a strongly
nonequilibrium stage of the inflationary expansion?
What was in the universe before the creation of vector bosons?
 Is it  possible to explain all visible matter
(with relict radiation and baryon-asymmetry)
by its cosmological creation from vacuum?
Why should the radiation-dominant stage in the Hot Universe Scenario
  again transfer into the inflationary stage,
 as it follows from the latest Supernova data~\cite{ps}?
 Is it possible, in principle, to explain  the
 appearance of the stage of an only radiation matter between two
 inflationary stages in the Hot Universe Scenario:
 fast (relict) and sluggish (Quintessence),
 both having no relationship to the visible matter of the Universe?

We have tried here, within the framework of the
Standard Model and General Relativity,
to answer these questions using
the relative measurement standard~\cite{weyl} and
the conformal cosmology~\cite{039}.
The stiff state in the framework of the
conformal cosmology describes
simultaneously all stages of the evolution of the universe:
the present-day stage of the accelerating evolution
(in the agreement
with the latest data on Supernova), the stage of primordial
element abundance in the universe, and the stage of creation
from geometrical vacuum of primordial
particles  most
singular in masses (in the Standard Model the only candidate for
a role of such primary particles is longitudinal vector
bosons $W, Z $). We have shown that
the temperature of the CMB radiation arises as an integral of
cosmic motion of the universe, which determines the initial data for the
Hubble constant
( $H_I\sim 2.7$K) and the running Planck mass ($ \vh_I\sim 10$~TeV).

During their lifetime vector bosons polarize the Dirac sea of
left fermions with negative energy.
The baryon asymmetry arises owing to three Sakharov
conditions: the
$ {\rm CP} $ - nonconservation, the evolution of the
universe and the violation of
the baryon number. From the ratio of the number of baryons to the number
of photons it is possible to make an estimation of the
value of the constant of a weak interaction.

The authors are grateful
to B.M. Barbashov and V.B. Priezzhev
for fruitful discussions. One of the authors (V.N.P.) is grateful to
B.A. Arbuzov for discussion of the problem of baryon asymmetry.
We are especially grateful to Prof. Y. Ne'eman for some important references to the
works on the conformal cosmology.

\end{document}